\documentclass[12pt]{article}
\usepackage{a4wide}
\usepackage{amssymb}
\usepackage{graphicx}
\begin{document}
{\renewcommand{\thefootnote}{\fnsymbol{footnote}}
\begin{center}
{\LARGE  Large effects from quantum reference frames}\\
\vspace{1.5em}
Martin Bojowald\footnote{e-mail address: {\tt bojowald@psu.edu}}
and Luis Mart\'{\i}nez
\\
\vspace{0.5em}
Institute for Gravitation and the Cosmos,\\
The Pennsylvania State
University,\\
104 Davey Lab, 251 Pollock Road, University Park, PA 16802, USA\\
\vspace{1.5em}
\end{center}
}

\setcounter{footnote}{0}

\begin{abstract}
  Reference frames are used to parameterize measurements of physical effects,
  but since their practical realization uses material objects, they may affect
  observations performed in a combined quantum state of the measured system
  together with the frame. Here, a procedure is used that makes it possible to
  describe non-monotonic reference scales in a quantum treatment, revealing
  large quantum effects in the measured system whenever a reference frame
  encounters a turning point.  Subtle quantum correlations in the combined
  state of system and frame, and more broadly the concept of relational
  quantum mechanics, can be tested via a characteristic and surprisingly large
  shift in the measured value.
\end{abstract}

\section{Introduction}

Direct measurement results inevitably refer to parameters, such as the
orientation in space or a time stamp, that do not correspond to physical
properties of the measured system but are nevertheless part of the outcome and
might change if the measurement is repeated. While such parameters are usually
treated as having real values of classical nature, a complete description
within an all encompassing quantum theory should take into account any quantum
properties they may have. More fundamentally, space and time may be subject to
quantization, just like matter, making it necessary to describe observations
by quantum reference frames \cite{QFR,RQM}.  Any given approach to such a
quantum theory of space-time or of gravity, provided it is sufficiently
tractable, then implies specific properties of quantum space and time that
would also apply to reference positions and directions in an experiment.
Corresponding physical effects are often expected to be tiny, but they may be
relevant for precision measurements or imply fundamental bounds on the
resolution that may be achieved, independently of what is well known from
uncertainty relations of quantum mechanics.

Based on these motivations, quantum reference frames have been analyzed in
various ways in recent years; see for instance \cite{QuantumRef, QuantumRef1,
  QuantumRefSpin, QuantumRef2, QuantumRef3, QuantumRef4, QuantumRef5,QuantumRefGroups,
  QuantumRef7}. Technical complications in most of these approaches often
require a restriction to reference variables on a monotonic scale, changing
uniformly from negative infinity to positive infinity. In practice, however,
reference directions must be restricted to bounded regions or may be wound up
in oscillatory fashion, such as the hands on an analog clock. Not much is known
about possible physical implications of reference variables with turning
points, except for an effect with oscillating fundamental clocks that, as
derived in \cite{LocalTime,Period} using methods of \cite{Gribov}, can be
surprisingly large, placing the lowest existing upper bound on the
possible period of a fundamental clock. Effects from turning points may be
large because they can affect the combined quantum state by changing phases,
superpositions, or even entanglement between system and frame. This conclusion
applies in particular to oscillating clocks with energy-dependent turning
points, while strictly periodic clocks with energy-independent turning points,
studied recently for instance in \cite{TimeTravel,PeriodicClocks},
in general have a weaker influence on possible measurements. 

The choice of clock variables can therefore have significant implications for
the physical behavior of a combination of clock and evolving system. In order
to analyze detailed properties of energy-dependent turning points, we
introduce here a model of a measurement by a quantized reference frame that
has only one energy-dependent turning point. The model is therefore neither
oscillating nor periodic, but it nevertheless shares crucial features with the
models used in \cite{LocalTime,Period}. (A cosmological version of a clock
with a single energy-dependent turning point has already been studied in
\cite{CosmoTurning}.) This feature makes it possible to derive analytical
results for properties of the system far from its turning point, while the
effects found in \cite{LocalTime,Period}, accumulated over several periods,
had to a large extent been seen using numerics. We find that even reference
frames with a single turning point reveal surprisingly large quantum effects
that may be measurable in model systems, providing a new opportunity to test a
fundamental feature of quantum reference frames in the laboratory. Instead of
trying to measure sensitive properties of an entangled state, the frame effect
makes it possible to test the implications indirectly through a characteristic
shift in one of the system observables. (This effect, already observed in the
cosmological model of \cite{CosmoTurning}, is reminiscent of discontinuous
jumps of observables derived in \cite{PeriodicClocks}. However, examples in the latter
paper used energy-independent turning points.)

Our specific system {with a single turning point of its
  reference frame} is not {strictly} designed to model a
clock, {although it can be used in this case in order to study
  a single half-cycle of an oscillating reference
  frame}. {More generally, it} represents a tractable but
non-monotonic quantum reference frame that may be realized in different ways,
{for instance by the relational distance between two particles
  in a trap}. We therefore obtain a novel non-trivial effect of quantum
reference frames and relational quantum mechanics, not necessarily related
{to but also applicable} to time and clocks.  We introduce our
model and relevant properties in Section~\ref{s:QRF}, followed by a discussion
of implications in Section~\ref{s:Impl}.

\section{Quantum reference frames with a linear clock potential}
\label{s:QRF}

Quantum reference frames {in the form discussed here} are
implemented by using the methods of constraints. {(See
  \cite{ReferenceFrames,RelativeSubsystems,OperationalReference,LocalReference}
  for alternative approaches.)}  If the reference observable is $\phi$, its
value changes by applying the translation operator
$\hat{p}_{\phi}=-i\hbar\partial/\partial\phi$ to a state {in
  the kinematical Hilbert space $L^2(\mathbb{R},{\rm
    d}\phi)\otimes {\cal H}$ with a system Hilbert space here assumed to be ${\cal
    H}=L^2(\mathbb{R},{\rm d}q)$, in order to be specific.} A system with basic
operators $(\hat{q},\hat{p})$ is observed relative to $\phi$ if it experiences
specific changes whenever $\phi$ changes. In general, the system change may
not be a simple translation of $q$ but rather some change implied by applying
an operator $H(\hat{q},\hat{p})$ to its state. The relational behavior of
$(q,p)$ relative to $\phi$ requires states $\psi$ with specific correlations,
implemented by the constraint equation
\begin{equation} \label{constraintnonrel}
  (\hat{p}_{\phi}+\hat{H})\psi=0\,.
\end{equation}
{In the $(\phi,q)$-representation of wave functions, solutions
  $\psi(\phi,q)$ of this first-order differential equation with normalizable
  initial data $\psi(\phi_0,q)\in{\cal H}$ at some $\phi_0$ imply a
  1-parameter family of states $\psi_{\phi}(q)=\psi(\phi,q)$ in the physical Hilbert space
  ${\cal H}$.}
If $\phi$ is time, equation~(\ref{constraintnonrel}) is equivalent to the Schr\"odinger equation
for $(q,p)$, but the concept of quantum reference frames is more general.

\subsection{Turning points}

Constraints are not required to be linear. For instance, relativistic
expressions are quadratic in energies and momenta, motivating a constraint
equation
\begin{equation}\label{constraintrel}
  (-\hat{p}_{\phi}^2+\hat{H}^2)\psi=0\,.
\end{equation}
This equation is not completely equivalent to the previous constraint because the
factorization
\begin{equation} \label{constraintfactor}
  (\hat{p}_{\phi}+\hat{H})(-\hat{p}_{\phi}+\hat{H})\psi=0
\end{equation}
(assuming that $\hat{H}$ does not depend on $\hat{\phi}$ and therefore
commutes with $\hat{p}_{\phi}$) leads to a new sign choice, depending on which
one of the two factors is used to act on the wave function. However, we 
do not gain new solutions if the original equation (\ref{constraintnonrel}) is
accompanied by its $\phi$-reversed version, mapping $\phi$ to
$-\phi$. {For $\phi$-independent $\hat{H}$, standard
  constructions of physical Hilbert spaces are available \cite{Blyth,GenRepIn}
  by restricting the Klein--Gordon bilinear form to suitable subsets of
  positive-norm solutions. This step is equivalent to adapting the physical
  Hilbert space obtained for (\ref{constraintnonrel}) by making a suitable
  replacement of $\hat{H}$ with $|\hat{H}|$, or by using the direct sum of
  the two Hilbert spaces obtained from $|\hat{H}|$ and $-|\hat{H}|$ if
  time-reversed solutions are of interest.}

New effects can be obtained if our assumption that $\hat{H}$ does not depend
on $\hat{\phi}$ is violated. If this is the case, $\hat{p}_{\phi}$ is not
conserved and may develop vanishing expectation values, corresponding to
classical turning points where $\phi$ changes direction. The constraint
equation together with reality or unitarity conditions can then limit the
range of possible values of $\langle\hat{\phi}\rangle$, modelling a bounded
range of a quantum reference frame. A $\phi$-dependent $\hat{H}$ also
complicates the analysis not only because it implies more challenging
differential equations, but also because the factorization
(\ref{constraintfactor}) is then inequivalent to the quadratic constraint
(\ref{constraintrel}). {Traditional methods to construct
  physical Hilbert spaces are no longer applicable in this case.} The approach
of \cite{Gribov}, together with further extensions given in \cite{Period},
makes it possible to deal with such models {while maintaining
  the system Hilbert space ${\cal H}$ as the physical Hilbert space, as we
  will demonstrate explicitly in Section~\ref{s:QuantumSolutions}.}

Specific examples of this form can be found in relativistic systems with
time-dependent potentials, using $p_{\phi}=E$ as the energy. The same behavior
can be modeled experimentally in graphene, which formally implies a
relativistic dispersion relation for electrons \cite{Graphene}. A
position-dependent effective mass then corresponds to a $\phi$-dependent
Hamiltonian in our model. As another example, while it is natural to interpret
quadratic terms in a Hamiltonian as kinetic energy, they may also appear in
specific potentials if a canonical transformation
$(\phi,p_{\phi};q,p)\mapsto (p_1,-x_1,p_2,-x_2)$ is used to swap configuration
and momentum coordinates. (The new variables have assigned roles as
configuration coordinates and momenta only according to their formal canonical
properties in Hamilton's equations, but not according to their units.) In this
way, one may model quadratic contributions $-p_{\phi}^2+p^2=-x_1^2+x_2^2$ to
our constraint as a combination of a harmonic and an inverted harmonic
potential for the two position coordinates $x_1$ and $x_2$ on a plane,
constrained by energy conservation. If kinetic energy is included,
$\frac{1}{2}p_1^2/m_1+\frac{1}{2}p_2^2/m_2=\frac{1}{2}\phi^2/m_1+\frac{1}{2}q^2/m_2$
implies a Hamiltonian of our general canonical form. In such systems, our
results do not refer to standard time evolution but rather to correlations
between the phase-space coordinates.

Relevant features are shown here by a specific system with an explicit
$\phi$-dependence chosen for mathematical tractability. Modeling a single
turning point of our reference variable $\phi$, we subject it to a linear
potential.  In order to make it easier to compare properties of states
relative to the turning point, we limit the potential to a bounded region,
writing it as $V(\phi)=\lambda \phi\theta(\phi)$ with Heaviside's step
function $\theta(\phi)$ and a positive constant $\lambda>0$. The classical
quadratic constraint (\ref{constraintrel}) is amended to
\begin{equation}
  C=-p_{\phi}^2-\lambda\phi \theta(\phi)+H(q,p)^2=0
\end{equation}
where $q$ and $p$ denote the canonical variables of a system measured with
respect to the frame $(\phi,p_{\phi})$. By requiring that $C=0$ for all
possible values of the phase-space variables, the constraint relates changes
$\delta q$ to changes $\delta\phi$, governed by Hamilton's equations generated
by $C$.  Since there are no direct coupling terms between the system $(q,p)$
and the frame $(\phi,p_{\phi})$, $H(q,p)$ and $p_{\phi}^2+V(\phi)$ are both
constant.  For any given value $H(q,p)=H$ on a specific solution, $\phi$
experiences a single turning point at $\phi_{\rm t}=H^2/\lambda$ where the
constraint implies $p_{\phi}=0$. The turning point depends on the energy $H$
of the evolving system. Since this energy is conserved, the dependence does
not have significant classical effects. However, there are characteristic
implications for quantum states that are superpositions of different system energies.

\subsection{Classical solutions}

The constraint describes correlations between $\phi$, $q$, and their
momenta. Instead of solving the constraint directly for one of the variables,
it is more convenient to determine how each of them depends on an auxiliary
parameter $\epsilon$ along the solutions of Hamilton's equations
\begin{equation} \label{phievolve}
  \frac{{\rm d}\phi}{{\rm d}\epsilon}=\frac{\partial C}{\partial p_{\phi}}=-2p_{\phi}\quad,\quad \frac{{\rm
      d}p_{\phi}}{{\rm d}\epsilon}=-\frac{\partial C}{\partial\phi}=\lambda \theta(\phi)\,.
\end{equation}
Once these equations as well as the corresponding ones for $q$ and $p$ have
been solved, inversions of the resulting $\epsilon$-dependent functions make
it possible to eliminate this parameter and derive the correlations as
functions $q(\phi)$ and $p(\phi)$ in which only the physical system and frame
variables appear. (See \cite{CosmoTurning} for a detailed discussion of gauge
parameters such as $\epsilon$.)

The required steps can be performed explicitly in our model, although the step
function makes the procedure more tedious than a simple linear potential would
suggest. Solving the second equation in (\ref{phievolve}), we obtain
\begin{equation}
  p_{\phi}(\epsilon)=\left\{\begin{array}{cl} c_1 &\mbox{if }\phi\leq 0\\
                              \lambda\epsilon+c_2 & \mbox{if
                                                    }\phi\geq 0 \end{array}\right.
\end{equation}
which then implies
\begin{equation}
  \phi(\epsilon)= \left\{\begin{array}{cl} -2c_1\epsilon+c_3 & \mbox{if
                                                               }\phi\leq 0\\
                           -\lambda\epsilon^2-2c_2\epsilon+c_4 & \mbox{if
                                                                 }\phi\geq0\end{array}\right.
                                                           \end{equation}
from the first equation. At this point, we have four integration constants
because we have solved two first-order differential equations, each one in the two
separate ranges of our piecewise linear potential. If we choose $\epsilon$
such that $\phi(0)=0$, which can be used as a boundary condition in both
ranges, we have $c_3=0=c_4$.
Continuity of $p_{\phi}(\epsilon)$ then implies $c_1=c_2$. The remaining constant,
$c_1=-H$, equals the negative conserved $H(q,p)$ upon using the constraint.

So far, our solutions are implicit because the conditions for different
branches refer to the sign of $\phi$, which changes along the solutions. With
our choice of boundary conditions, they can be turned into explicit solutions
by rearranging the branches according to
\begin{equation}
  \phi(\epsilon) = \left\{\begin{array}{cl} 2H\epsilon & \mbox{if }
                                                            \epsilon\leq0\\
                              -\lambda\epsilon^2+2H\epsilon & \mbox{if
                              }0\leq\epsilon\leq 2H/\lambda\\ -2H\epsilon+4
                              H^2/\lambda & \mbox{if }
                                            \epsilon\geq 2H/\lambda \end{array}\right.
\end{equation}
and
\begin{equation}
p_{\phi}(\epsilon) = \left\{\begin{array}{cl} -H & \mbox{if }\epsilon\leq 0\\
                                \lambda\epsilon-H & \mbox{if
                                                    }0\leq \epsilon\leq 2H/\lambda\\ H
                                                   & \mbox{if
                                                     }\epsilon\geq 2H/\lambda \end{array}\right. \,.
\end{equation}

These solutions are valid for any $H(q,p)$, provided the $\phi$-potential is
of the given form. To proceed further, we have to choose a specific system
Hamiltonian, for which we use the simplest non-trivial case, $H=p$, in order to
outline the general procedure. Without the turning point, $q(\phi)$ would
be linear, but as we will see, this behavior is modified around a
turning point. Again using Hamilton's equations generated by $C$, now applied
to $q$ and $p$, $p(\epsilon)$ is constant and we have
$q(\epsilon)=2p\epsilon+q_0$ as a function of the auxiliary parameter
$\epsilon$. It is therefore straightforward to invert $q(\epsilon)$ for $\epsilon(q)$,
insert the result in $\phi(\epsilon)$, and
find the $\epsilon$-independent relation
\begin{equation}
  \phi(q)=\left\{\begin{array}{cl} q-q_0 & \mbox{if }q\leq q_0\\
                   -\frac{1}{4}\lambda (q-q_0)^2/p^2+q-q_0 & \mbox{if
                                                             }q_0\leq q\leq q_0+4p^2/\lambda\\
                   q_0-q+4p^2/\lambda & \mbox{if
                                        }q\geq q_0+4p^2/\lambda\end{array}\right. \,.
\end{equation}

The physical interpretation of our model requires another inversion because we
use $\phi$ as the reference frame with a turning point, on which $q$ depends
as determined by $q(\phi)$.  A local inversion for $q(\phi)$ in the three
branches, combined with continuity in $\phi$, then implies
\begin{equation}
  q(\phi)= \left\{\begin{array}{cl} q_0+\phi & \mbox{if }\phi\leq 0\\
                    q_0+2p^2\lambda^{-1} (1-\sqrt{1-\lambda\phi/p^2}) &
                                                                        \mbox{if
                                                                        }0\leq
                                                                        \phi\leq
                                                                        p^2/\lambda\\
                    q_0+2p^2\lambda^{-1}(1+\sqrt{1-\lambda\phi/p^2}) &
                                                                       \mbox{if
                                                                       }p^2/\lambda\geq
                                                                       \phi\geq
                                                                       0\\
                    q_0-\phi+4p^2/\lambda & \mbox{if }\phi\leq 0 \end{array}\right.\,.
\end{equation}
The first two lines describe measurements before the turning points, and the
last two lines after.

The combined equation describes a double-valued function $q(\phi)$ because the
original variable $\phi$ does not distinguish between values before and after
the turning point. In order to keep track of when the turning point has been
encountered, we introduce an effective monotonic scale $\tau$ constructed from
the frame variable $\phi$. This method is related to unwinding oscillating or
periodic clocks, as used in
\cite{DiracChaos,DiracChaos2,LocalTime,Period,PeriodicClocks}, but here we
have only two branches rather than an infinite number of cycles.

Before the turning point is reached, a simple identification $\phi=\tau$ for
$\tau<\phi_{\rm t}$ obeys the required condition of monotonic
$\tau$-dependence for all system variables. After the turning point, $\tau$
should continue to increase while $\phi$ decreases, such that $\phi=-\tau+c$
with a constant $c$.  Continuity then uniquely determines
\begin{equation} \label{phitau}
  \phi(\tau)=\left\{\begin{array}{cl} \tau & \mbox{if }\tau\leq p^2/\lambda\\
                      -\tau+2p^2/\lambda & \mbox{if
                                           }\tau\geq p^2/\lambda\end{array}\right.\,.
\end{equation}
                                     
Finally, classical relational evolution with respect to the constructed
monotonic scale $\tau$ is given by
\begin{equation} \label{qtau}
  q(\tau) = \left\{\begin{array}{cl} q_0+\tau & \mbox{if }\tau<0\\
                    q_0+2p^2\lambda^{-1} (1-\sqrt{1-\lambda\tau/p^2}) &
                                                                        \mbox{if
                                                                        }0\leq
                                                                        \tau\leq
                                                                        p^2/\lambda\\
                    q_0+2p^2\lambda^{-1}(1+\sqrt{\lambda\tau/p^2-1}) &
                                                                       \mbox{if
                                                                       }p^2/\lambda\leq
                                                                       \tau\leq
                                                                       2p^2/\lambda\\
                    q_0+\tau+2p^2/\lambda & \mbox{if }\tau\geq 2p^2/\lambda \end{array}\right.\,.
\end{equation}
Extrapolating the final $q(\tau)=q_0+\tau+2p^2/\lambda$ back to an initial $\tau=0$, the position
appears increased by an amount
\begin{equation}
  \delta q_{\rm class}=\frac{2p^2}{\lambda}
\end{equation}
compared with the result obtained from $q(\tau)=q_0+\tau$.  This shift happens
because the position $q$ has been changing at a constant rate with respect to
$\epsilon$ while the frame variable $\phi$ was being slowed down and reversed
by the linear potential. Such a displacement can serve as an observable
parameter that does not require direct access to the turning point. We will
now turn to the quantum theory of this system and derive the corresponding
displacement.

\subsection{Quantum solutions}
\label{s:QuantumSolutions}

Quantum measurements relative to $\phi$ can be formulated by solving the
classical constraint for $p_{\phi}=\pm \sqrt{p^2-V(\phi)}$ and quantizing the
two sides:
\begin{equation} \label{Schroedingerphi}
  i\hbar \frac{\partial\psi(\phi,p)}{\partial\phi} =\pm
  \sqrt{p^2-\lambda\phi\theta(\phi)} \;\psi(\phi,p)\,.
\end{equation}
The sign choice indicates that this equation determines independent solutions
for the two branches, before and after the turning point. But neither of them
provide unitary changes of $q$ with respect to a global coordinate $\phi$
because the square root is not guaranteed to be real for all
$\phi$. Physically, unitary changes for any $\phi$ would in fact be
problematic because we know that $\phi$ has a turning point and cannot take
all real values.

Instead of using the non-monotonic frame variable $\phi$, we should transform
the wave equation to the effective monotonic scale $\tau$ from (\ref{phitau}), which can take
all real values in the classical theory. Moreover, unitary changes relative to
$\tau$ would unambiguously connect the two branches before and after the
turning point. Since $\phi$ is directly related to $\tau$, we can use the
chain rule to transform the $\phi$-equation (\ref{Schroedingerphi}) to a
$\tau$-equation:
\begin{equation} \label{Schroedingertau}
  i\hbar \frac{\partial\psi(\tau,p)}{\partial\tau} =\pm \frac{{\rm
      d}\phi}{{\rm d}\tau}
  \sqrt{p^2-\lambda\phi(\tau)\theta(\phi(\tau))} \psi(\tau,p)\,.
\end{equation}
Because $\sqrt{p^2-V(\phi(\tau))}$ is always real, by construction of $\tau$,
changes relative to $\tau$ are unitary.   The sign ambiguity can be resolved by
requiring a positive Hamiltonian for forward $\tau$-changes.  Therefore, we
choose the plus sign in front of the square root while
${\rm d}\phi/{\rm d}\tau=1$ (before the turning point) and the minus sign
while ${\rm d}\phi/{\rm d}\tau=-1$ (after the turning point).

{As a consequence of unitarity, solutions of
  (\ref{Schroedingertau}) with normalizable initial date
  $\psi(\tau_0,p)\in{\cal H}$ at some fixed $\tau_0$ imply 1-parameter
  families $\psi_{\tau}(p)=\psi(\tau,p)$ of states in the physical Hilbert
  space ${\cal H}$. Our constructions therefore solve the Hilbert-space
  problem of traditional approaches and are applicable to relational
  measurements with $\phi$-dependent Hamiltonians.}

In order to obtain explicit solutions, we go back to the $\phi$-equation
(\ref{Schroedingerphi}) and solve it in the $p$-representation of wave
functions. The general solution takes the form
$\psi(\phi,p)=f(p) \exp(\mp i\hbar^{-1} \Theta(\phi,p))$ with an arbitrary
function $f(p)$ (determined by an initial state) and the phase
\begin{equation} \label{Theta}
  \Theta(\phi,p) = \left\{\begin{array}{cl} p\phi & \mbox{if }\phi\leq 0\\
                            -\frac{2}{3}\lambda^{-1}
                            (p^2-\lambda\phi)^{3/2}+\frac{2}{3}\lambda^{-1}
                            p^3 & \mbox{if
                                  }0\leq \phi\leq p^2/\lambda\end{array}\right.
\end{equation}
using continuity in $\phi$.  Global dependence on the monotonic $\tau$, via
$\phi(\tau)$ in (\ref{phitau}), is given by the wave function
\begin{equation}\label{psitaup}
  \psi(\tau,p) = \left\{\begin{array}{cl} f(p)\exp(-i\hbar^{-1}
                            \Theta(\phi(\tau),p) & \mbox{if
                                                   }\tau\leq p^2/\lambda\\ f(p)
                            \exp(i\hbar^{-1}(\Theta(\phi(\tau),p)-\frac{4}{3}p^3/\lambda))
                                                 & \mbox{if
                                                   }\tau\geq p^2/\lambda\end{array}\right.
\end{equation}
in which we already implemented the unique sign choice of the phase in
(\ref{Schroedingertau}). The phase shift in the second line is required by
continuity of the wave function.  We have coefficients
$f(p)=\psi(\tau_0,p)\exp(i\hbar^{-1}p\tau_0)$ determined by the state at some
$\tau_0\leq 0$, at which none of the $p$-contributions have yet reached their
turning point. Using the solution (\ref{Theta}), we have
\begin{equation} \label{psi}
  \psi(\tau,p)= \left\{\begin{array}{cl} f(p)\exp(-i\hbar^{-1}p\tau) & \mbox{if
                                                                   }\tau\leq 0\\
                         f(p)\exp(-\frac{2}{3}i\hbar^{-1}\lambda^{-1}
                         (p^3-(p^2-\lambda\tau)^{3/2})) & \mbox{if
                                                         }0\leq \tau\leq p^2/\lambda\\
                         f(p)\exp(-\frac{2}{3}i\hbar^{-1}\lambda^{-1}(p^3+(\lambda\tau-p^2)^{3/2}))
                                                                     &
                                                                       \mbox{if
                                                                       }p^2/\lambda\leq
                                                                       \tau\leq
                                                                       2p^2/\lambda\\
                         f(p)\exp(-i\hbar^{-1}p\tau+\frac{2}{3}
                         i\hbar^{-1}\lambda^{-1} p^3) & \mbox{if
                                                        }\tau\geq 2p^2/\lambda \end{array}\right.\,.
\end{equation}
The position expectation value therefore equals
\begin{eqnarray} \label{q}
  \langle\hat{q}\rangle(\tau) &=& i\hbar\langle
                                  \psi(\tau,p),\partial\psi(\tau,p)/\partial p\rangle\nonumber\\
  &=&
                                  \left\{\begin{array}{cl} q_0+\tau & \mbox{if
                                                                      }\tau\leq
                                                                      0\\
                                           q_0+2\lambda^{-1}\left\langle \hat{p}^2-\hat{p}
                                           \sqrt{\hat{p}^2-\lambda\tau}\right\rangle & \mbox{if
                                                                     }0\leq
                                                                                 \tau\leq p^2/\lambda\\
                                           q_0+2\lambda^{-1}\left\langle
                                           \hat{p}^2-\hat{p}\sqrt{\lambda\tau-\hat{p}^2}\right\rangle 
                                                                    & \mbox{if
                                                                      }p^2/\lambda\leq
                                                                      \tau\leq
                                                                      2p^2/\lambda\\
                                           q_0+\tau-2\lambda^{-1}\langle \hat{p}^2\rangle &
                                                                       \mbox{if
                                                                       }\tau\geq
                                                                                      2p^2/\lambda
                                         \end{array}\right.\nonumber
\end{eqnarray}
with $q_0=i\hbar \langle f(p),\partial f(p)/\partial p\rangle$. The
expectation values remaining in this expression can be taken in the
state $f(p)$, using $\tau$ as a parameter rather than an integration variable in the
inner product.

\begin{figure}
    \centering
    \includegraphics[width=0.8\textwidth]{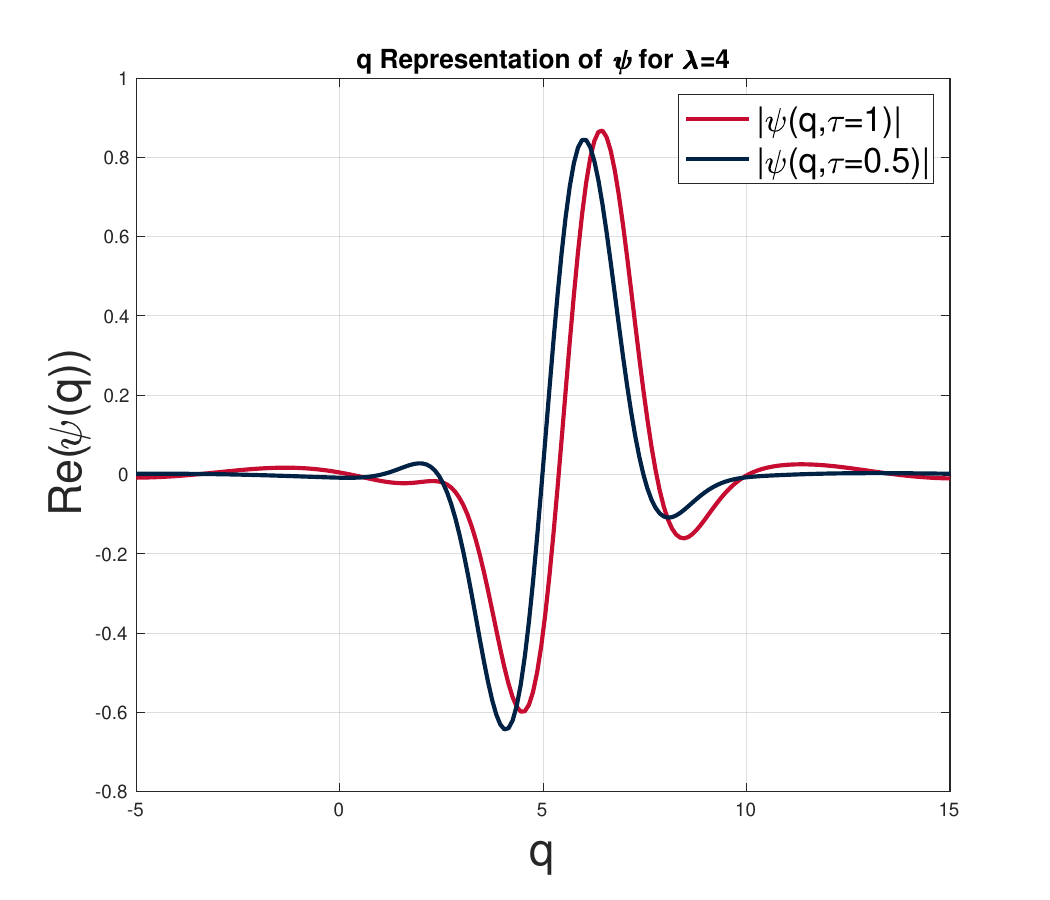}
    \caption{{A quantum state evolving relative to a reference
        frame with a turning point experiences additional phase changes
        compared with a quantum state evolving relative to a global frame. For
        our main example of $\hat{H}=\hat{p}$, global evolution would rigidly
        translate the wave function. The wave functions shown here shows more
        subtle changes when it refers to a quantum state evolving relative to
        a reference frame with a turning point. In this example, most of the momentum
        contributions in its eigenstate decomposition encounter the turning
        point in the time period between $\tau=0.5$ and $\tau=1$ used
        here. The time-dependent position expectation value of the same state
        is shown in Fig.~\ref{Fig:Shift}, which demonstrates that crucial
        turning-point effects happen in this time period.}
      \label{Fig:Phase}}
\end{figure}

\section{Implications}
\label{s:Impl}

Our prime observable of interest is the displacement of the position at
$\tau=0$ obtained by extrapolating the asymptotic
$\langle\hat{q}\rangle(\tau)$ for large $\tau$ back to $\tau=0$:
\begin{equation}
  \delta q_{\rm quantum}=-\frac{2\langle\hat{p}^2\rangle}{\lambda}\,.
\end{equation}
The sign here is the opposite of what we found in the classical displacement.
The overall shift between the classical and quantum behavior after the turning
point is therefore significant, given by
\begin{equation} \label{delta}
  \delta q= \delta q_{\rm quantum}-\delta q_{\rm classical}=-\frac{4\langle \hat{p}\rangle^2}{\lambda}- \frac{2(\Delta
    p)^2}{\lambda}\,. 
\end{equation}
The second term looks like a typical semiclassical quantum correction,
depending on momentum fluctuations. (Here, we interpret the classical $p$ as the
expectation value $\langle\hat{p}\rangle$. If we interpret the classical $p^2$
as the expectation value $\langle\hat{p}^2\rangle$, the fluctuation term will
have a different coefficient.) The first term, however, does not depend
on $\hbar$ and therefore constitutes an unexpectedly large quantum effect.

The main contribution to the shift is not a traditional semiclassical
result. It rather depends on properties of the phase while the state travels
through the turning point. Since the phase is always multiplied by
$\hbar^{-1}$ in the wave function, it is possible for $\hbar$ to cancel out in
suitable observables, just as it happens in our specific example.  Phase
effects are often counter-intuitive, and our result is no exception.
Formally, we observe that the phase receives characteristic features from the
presence of a turning point, not only because $\Theta$ is non-linear in $\tau$
around the turning point (a standard implication of a $\phi$-dependent
Hamiltonian such as $(p^2-V(\phi))^{1/2}$) but also because different
$p$-contributions to $\psi(\tau,p)$ experience their turning points at
different $\phi$ or $\tau$. The state therefore does not behave as a simple translation
$H=p$ would suggest; see Fig.~\ref{Fig:Phase}.

Continuity of the phase through different branches around the turning point
leads to the specific $p$-dependence shown in (\ref{psi}). In this momentum
representation, changes of $\langle\hat{q}\rangle$ are encoded in the
$p$-dependence of the phase, which remains bounded and slowly-varying around
turning points. This behavior is in contrast to the classical $q(\phi)$ which
changes rapidly around a turning point of $\phi$ because
${\rm d}q/{\rm d}\phi=({\rm d}q/{\rm d}\epsilon)/({\rm d}\phi/{\rm
  d}\epsilon)=-p/p_{\phi}$ is large and diverges at $p_{\phi}=0$. The quantum
state implies a similar term proportional to $p_{\phi}^{-1}$ in
$\langle\hat{q}^2\rangle$, resulting from the second derivative of the phase
in the momentum representation. Position fluctuations are therefore large
around an energy-dependent turning point, which is expected because different
energy eigenstates in a superposition reach their turning points at different
times, spreading out the state. The rate of change
${\rm d}\langle\hat{q}\rangle/{\rm d}\tau$ contains the classically diverging
term, but only for one value of $p$ at any given $\tau$. The classical
divergence is then averaged out in the quantum expectation value.

\begin{figure}
    \centering
    \includegraphics[width=0.9\textwidth]{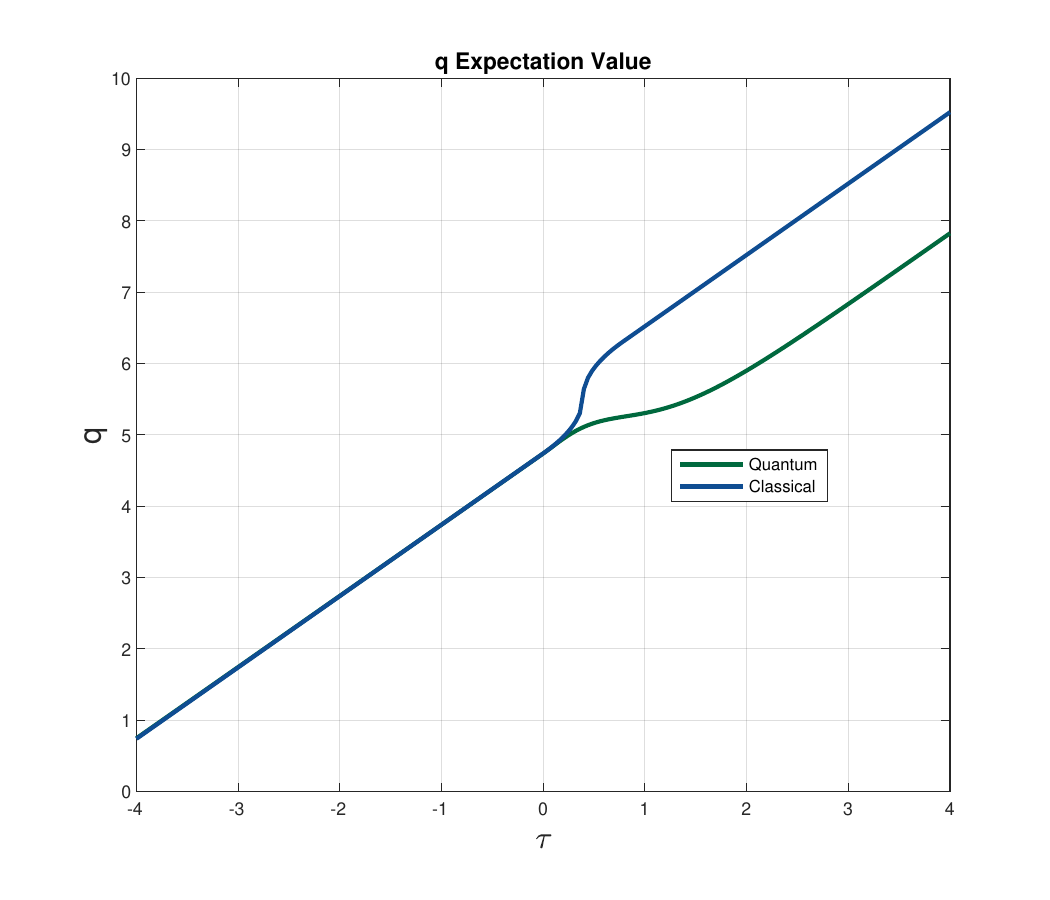}
    \caption{Classical solution and expectation value for the position in a
      numerically solved quantum state, both as a function of
      $\tau$. The quantum calculation is based on a Gaussian state
      centered at $q_0=4$ with mean momentum $p_0=1.25$ and variance
      $\sigma=1$. The linear potential of the frame variable $\phi$ has a
      slope $\lambda=4$. With these values,
      $(\Delta p)^2=\frac{1}{4}\sigma^{-2}= 0.25$ and the shift $\delta q$ in
      (\ref{delta}) evaluates to $-1.6875$ in agreement with the vertical scale.  \label{Fig:Shift}}
\end{figure}

Properties
of the phase lead to a much smaller quantum change in the turning-point region
according to (\ref{q}), illustrated in Fig.~\ref{Fig:Shift}. Our result is
therefore a direct consequence of the quantum treatment of energy-dependent
turning points: A turning point is necessary for the classical rate of change
to be large, such that its quantum suppression can be
significant. Energy-dependence of the turning point is implied by the
condition that $p_{\phi}$ cannot be conserved if $p_{\phi}=0$ is to be reached
along classical solutions, which requires an explicit $\phi$-dependence in the
clock Hamiltonian. Solving the constraint for turning-point values at
$p_{\phi}=0$ then leads to turning-point $\phi$ depending on the system energy
$H$.

These relational properties with respect to a local frame with a turning point
conspire to produce a large quantum correction in the asymptotic
shift. {Estimates for potential experimental tests depend on
  the precise realization and the system Hamiltonian. Qualitatively, we may
  assume that the linear potential for the frame degree of freedom $\phi$ is
  implied by the gravitational field on Earth, such that $\lambda=m^2g$ with
  the gravitational acceleration $g$ and the mass $m$ of the corresponding
  particle with vertical position $\phi$. (In this equation for $\lambda$, the
  mass is squared since we did not include an inverse-mass factor in our
  momentum term.) In an ensemble of atoms of mass $m$, the momentum
  expectation value can be related to the temperature $T$ by
  $\langle\hat{p}\rangle^2/m\sim k_{\rm B}T$. Thus,
  \begin{equation}
    \frac{\delta q}{{\rm m}}\sim \frac{k_{\rm
    B}T}{mg{\rm m}}\sim 10^{-24} \frac{{\rm kg}}{m} \frac{T}{{\rm K}} \sim 10^3 \frac{{\rm
    amu}}{m} \frac{T}{{\rm K}} 
\end{equation}
by orders of magnitude. For an atom of mass about $100{\rm amu}$, this is a
large distance of $\delta q\sim 10{\rm m}$ at $T\sim 1{\rm K}$, but
controlling the frame-system atom pair in a coherent state requires much lower
temperatures. Currently, atom traps operating at $T\sim 10^{-6}{\rm K}$ are in
reach \cite{MicroKelvinTrap}, which implies a quantum shift
of $\delta q\sim 10^{-5} {\rm m}$. In addition to the shift, the time required to
move through the turning point is important because it is limited in an
experiment by the ability to maintain quantum coherence of the frame-system pair. This
time can be estimated as
\begin{equation}
\frac{\delta\tau}{{\rm s}}\sim \frac{m\delta q}{\langle\hat{p}\rangle {\rm s}} \sim \frac{\sqrt{k_{\rm
      B}T}}{\sqrt{m}g {\rm s}}\sim 10\sqrt{\frac{T}{{\rm K}}\frac{\rm
    amu}{m}}\,,
\end{equation}
giving a large value of $\delta \tau\sim 1{\rm ms}$ at $T\sim 10^{-6}{\rm K}$
for $m\sim 100{\rm amu}$. The required coherence time can be reduced by
replacing the gravitational force with a large electric force acting on an ion,
at the expense of decreasing the expected quantum shift $\delta q$.
} 

The qualitative features {demonstrated here} are independent
of the specific frame potential and system Hamiltonian used for our model.
Analytical methods are still available for a general system Hamiltonian
$H(q,p)$ if we replace our $p$ here with the energy spectrum and work with
states in the $\hat{H}$-representation $\psi(\tau,E)$.
{Equations~(\ref{Schroedingerphi})--(\ref{psi}) remain valid
  with a simple substitution of $E$ for $p$. However, the energy operator
  $\hat{H}$ may not have a strict canonically conjugate operator that could be
  quantized to $i\hbar\partial/\partial E$ as a substitute of $\hat{q}$ in the
  $p$-representation used in (\ref{q}). In general, such a system time
  operator only exists as a positive operator-valued measure, which could be
  applied in this case in combination with related constructions in
  \cite{QuantumRef, QuantumRef1, QuantumRefSpin, QuantumRef2, QuantumRef3,
    QuantumRef4, QuantumRef5,QuantumRefGroups, QuantumRef7}. Alternatively,
  the quantum shift derived here does not require a strict conjugate operator
  but can be explored for on the operator level using an accessible observable
  that does not commute with $\hat{H}$.}
Some realistic choices
{of energy operators} may be hard to solve analytically
{and it may be challenging to find suitable superpositions of
  energy eigenstates for a desired initial state such as a Gaussian,} but
numerics as used for our figures would still be available. If quantum
reference frames with a turning point can be realized in an experimental
setup, it would be possible to test implications by looking for characteristic
shifts of system observables even if direct access to the correlated state and
its phase may not be possible.

\medskip

\section*{Acknowledgements}
This work was supported in part by the Sloan
Foundation and Penn State's Office for Educational Equity, and by NSF grant PHY-2206591.


\begin{thebibliography}{10}

\bibitem{QFR}
Y.\ Aharonov and T.\ Kaufherr,
\newblock Quantum frames of reference,
\newblock {\em Phys.\ Rev.\ D} 30 (1984) 368--385

\bibitem{RQM}
C.\ Rovelli,
\newblock Relational Quantum Mechanics,
\newblock {\em Int.\ J.\ Theor.\ Phys.} 35 (1996) 1637, [quant-ph/9609002]

\bibitem{QuantumRef}
E.\ Castro-Ruiz, F.\ Giacomini, and C.\ Brukner,
\newblock Entanglement of quantum clocks through gravity,
\newblock {\em PNAS} 114 (2017) E2303, [arXiv:1507.01955]

\bibitem{QuantumRef1}
F.\ Giacomini, E.\ Castro-Ruiz, and C.\ Brukner,
\newblock Quantum mechanics and the covariance of physical laws in quantum
  reference frames,
\newblock {\em Nat.\ Commun.} 10 (2019) 494, [arXiv:1712.07207]

\bibitem{QuantumRefSpin}
F.\ Giacomini, E.\ Castro-Ruiz, and C.\ Brukner,
\newblock Relativistic quantum reference frames: the operational meaning of
  spin,
\newblock {\em Phys.\ Rev.\ Lett.} 123 (2019) 090404, [arXiv:1811.08228]

\bibitem{QuantumRef2}
A.\ Vanrietvelde, P.~A.\ Hoehn, F.\ Giacomini, and E.\ Castro-Ruiz,
\newblock A change of perspective: switching quantum reference frames via a
  perspective-neutral framework,
\newblock {\em Quantum} 4 (2020) 225, [arXiv:1809.00556]

\bibitem{QuantumRef3}
A.\ Vanrietvelde, P.~A.\ Hoehn, and F.\ Giacomini,
\newblock Switching quantum reference frames in the $N$-body problem and the
  absence of global relational perspectives, [arXiv:1809.05093]

\bibitem{QuantumRef4}
P.~A.\ Hoehn, A.~R.~H.\ Smith, and M.~P.~E.\ Lock,
\newblock The Trinity of Relational Quantum Dynamics,
\newblock {\em Phys.\ Rev.\ D} 104 (2021) 066001, [arXiv:1912.00033]

\bibitem{QuantumRef5}
P.~A.\ Hoehn, A.~R.~H.\ Smith, and M.~P.~E.\ Lock,
\newblock Equivalence of approaches to relational quantum dynamics in
  relativistic settings,
\newblock {\em Front.\ Phys.} 9 (2021) 587083, [arXiv:2007.00580]

\bibitem{QuantumRefGroups}
A.-C.\ de~la Hamette, T.~D.\ Galley, P.~A.\ Hoehn, L.\ Loveridge, and M.~P.\
  Mueller,
\newblock Perspective-neutral approach to quantum frame covariance for general
  symmetry groups, [arXiv:2110.13824]

\bibitem{QuantumRef7}
F.\ Giacomini and A.\ Kempf,
\newblock Second-quantized Unruh-DeWitt detectors and their quantum reference
  frame transformations,
\newblock {\em Phys.\ Rev.\ D} 105 (2022) 125001, [arXiv:2201.03120]

\bibitem{LocalTime}
G.\ Wendel, L.\ Mart\'{\i}nez, and M.~Bojowald,
\newblock Physical implications of a fundamental period of time,
\newblock {\em Phys.\ Rev.\ Lett.} 124 (2020) 241301, [arXiv:2005.11572]

\bibitem{Period}
M.\ Bojowald, L.\ Mart\'{\i}nez, and G.\ Wendel,
\newblock Relational evolution with oscillating clocks,
\newblock {\em Phys.\ Rev.\ D} 105 (2022) 106020, [arXiv:2110.07702]

\bibitem{Gribov}
M.~M.\ Amaral and M.\ Bojowald,
\newblock A path-integral approach to the problem of time,
\newblock {\em Ann.\ Phys.} 388C (2018) 241--266, [arXiv:1601.07477]

\bibitem{TimeTravel}
A.\ Alonso-Serrano, S.\ Schuster, and M.\ Visser,
\newblock Emergent Time and Time Travel in Quantum Physics,
\newblock {\em Universe} 10 (2024) 73, [arXiv:2312.05202]

\bibitem{PeriodicClocks}
L.\ Chataignier, P.~A.\ H\"ohn, M.~P.~E.\ Lock, and F.~M.\ Mele,
\newblock Relational Dynamics with Periodic Clocks, [arXiv:2409.06479]

\bibitem{CosmoTurning}
L.\ Mart\'{\i}nez, M.\ Bojowald, and G.\ Wendel,
\newblock Freeze-free cosmological evolution with a non-monotonic internal
  clock,
\newblock {\em Phys.\ Rev.\ D} 108 (2023) 086001, [arXiv:2309.07825]

\bibitem{ReferenceFrames}
S.~D.\ Bartlett, T.\ Rudolph, and R.~W.\ Spekkens,
\newblock Reference frames, superselection rules, and quantum information,
\newblock {\em Rev.\ Mod.\ Phys.} 79 (2007) 555--606, [arXiv:quant-ph/0610030]

\bibitem{RelativeSubsystems}
E.\ Castro-Ruiz and O.\ Oreshkov,
\newblock Relative subsystems and quantum reference frame transformations,
  [arXiv:2110.13199]

\bibitem{OperationalReference}
T.\ Carette, J.\ Glowacki, and L.\ Loveridge,
\newblock Operational quantum reference frame transformations,
\newblock {\em Quantum} 9 (2025) 1680, [arXiv:2303.14002]

\bibitem{LocalReference}
C.~J.\ Fewster, D.~W.\ Janssen, L.~D.\ Loveridge, K.\ Rejzner, and J.\ Waldron,
\newblock Quantum reference frames, measurement schemes and the type of local
  algebras in quantum field theory,
\newblock {\em Commun.\ Math.\ Phys.} 406 (2025) 19, [arXiv:2403.11973]

\bibitem{Blyth}
W.~F.\ Blyth and C.~J.\ Isham,
\newblock Quantization of a Friedmann universe filled with a scalar field,
\newblock {\em Phys.\ Rev.\ D} 11 (1975) 768--778

\bibitem{GenRepIn}
J.~B.\ Hartle and D.\ Marolf,
\newblock Comparing Formulations of Generalized Quantum Mechanics for
  Reparametrization-Invariant Systems,
\newblock {\em Phys.\ Rev.\ D} 56 (1997) 6247--6257, [gr-qc/9703021]

\bibitem{Graphene}
A.~H.\ Castro~Neto, F.\ Guinea, N.~M.~R.\ Peres, K.~S.\ Novoselov, and A.~K.\
  Geim,
\newblock The electronic properties of graphene,
\newblock {\em Rev.\ Mod.\ Phys.} 81 (2009), [arXiv:0709.1163]

\bibitem{DiracChaos}
B.\ Dittrich, P.~A.\ Hoehn, T.~A.\ Koslowski, and M.~I.\ Nelson,
\newblock Chaos, Dirac observables and constraint quantization,
  [arXiv:1508.01947]

\bibitem{DiracChaos2}
B.\ Dittrich, P.~A.\ Hoehn, T.~A.\ Koslowski, and M.~I.\ Nelson,
\newblock Can chaos be observed in quantum gravity?,
\newblock {\em Phys.\ Lett.\ B} 769 (2017) 554--560, [arXiv:1602.03237]

\bibitem{MicroKelvinTrap}
A.~R.\ Ferdinand, A.\ Luo, S.\ Jammi, Z.\ Newman, G.\ Spektor, O.\ Koksal,
  P.~B.\ Patel, D.\ Sheredy, W.\ Lunden, A.\ Rakholia, T.~C.\ Briles, W.\ Zhu,
  M.~M.\ Boyd, A.\ Agrawal, and S.~B.\ Papp,
\newblock Laser cooling ${}^{88}$Sr to microkelvin temperature with an
  integrated-photonics system,
\newblock {\em Phys.\ Rev.\ Applied} 23 (2025) L031002, [arXiv:2404.13210]

\end{thebibliography}

\end{document}